\documentclass[12pt]{iopart}
\usepackage{color}
\usepackage{epic}
\usepackage{iopams}
\usepackage{setstack}
\usepackage{eucal}
\usepackage{accents}

\eqnobysec

\newcommand{\wh}{\widehat}
\newcommand{\wt}{\widetilde}
\newcommand{\wb}{\widebar}
\DeclareMathAccent{\wtilde}{\mathord}{largesymbols}{"65}
\DeclareMathAccent{\what}{\mathord}{largesymbols}{"62}
\DeclareMathAccent{\wbar}{\mathord}{largesymbols}{"63}

\makeatletter
\def\wb{\accentset{{\cc@style\underline{\mskip10mu}}}}
\makeatother

\def\m@th{\mathsurround=0pt}
\mathchardef\bracell="0365
\def\upbrall{$\m@th\bracell$}
\def\undertilde#1{\mathop{\vtop{\ialign{##\crcr
 $\hfil\displaystyle{#1}\hfil$\crcr
 \noalign
 {\kern1.5pt\nointerlineskip}
 \upbrall\crcr\noalign{\kern1pt
 }}}}\limits}

\newcommand{\gm}{\gamma}

\newcommand{\ld}{\lambda}

\newcommand{\be}{\begin{equation}}
\newcommand{\ee}{\end{equation}}
\newcommand{\bea}{\begin{eqnarray}}
\newcommand{\eea}{\end{eqnarray}}
\newcommand{\bse}{\begin{subequations}}
\newcommand{\ese}{\end{subequations}}
\newcommand{\nn}{\nonumber}

\newcommand{\ut}{\undertilde}

\newcommand{\ol}{\overline}

\newcommand{\bL}{\boldsymbol{L}}

\newcommand{\bM}{\boldsymbol{M}}

\begin{document}
\title[Chazy-type difference equations]{$q$-difference equations of KdV type
and \lq\lq Chazy-type\rq\rq\ second-degree difference equations} 

\author{Chris M Field}
\address{Korteweg-de Vries Institute for Mathematics, Universiteit van Amsterdam,
Plantage Muidergracht 24,
1018 TV Amsterdam,
The Netherlands. E-mail: cfield@science.uva.nl}
\author{Nalini Joshi}
\address{School of Mathematics and Statistics F07, University of Sydney, NSW 2006 Australia. E-mail:  nalini@maths.usyd.edu.au}
\author{Frank W Nijhoff}
\address{School of Mathematics, University of Leeds, Leeds LS2 9JT, UK. E-mail: nijhoff@maths.leeds.ac.uk }
\date{\today}
\begin{abstract}
By imposing special compatible similarity constraints on a class of integrable partial $q$-difference equations of KdV-type we derive a hierarchy of second-degree 
ordinary $q$-difference equations. 
The lowest (non-trivial)
member of this hierarchy is a second-order second-degree equation which can be considered as an analogue of equations in the class studied by Chazy. We present corresponding isomonodromic deformation problems and discuss the relation between this class of difference equations and other equations of Painlev\'e type. 


\end{abstract}

\section{Introduction}
\setcounter{equation}{0}

The construction and study of discrete Painlev\'e equations has been a topic of research interest 
for almost two decades, \cite{NP,FIK,RGH,JimSakai}. Reviews of the subject may be found in \cite{GNR,GR}.
The subject has culminated in the classification by H. Sakai of discrete as well continuous 
Painlev\'e equations based on the algebraic geometry of the corresponding rational 
surfaces associated with the spaces of initial conditions \cite{Sakai}. 
As a byproduct of the latter treatment, 
a ``mistress'' discrete Painlev\'e equation with elliptic dependence on the independent variable
was discovered.  

In the history of the Painlev\'e program, after the classification results for second-order first-degree equations, Painlev\'e's students, Chazy and Garnier, \cite{Chazy9,Chazy11,Garn}, investigated the classification of second-order second-degree equations and third-order first-degree equations. The classification of the second-degree class was completed by Cosgrove in recent years, \cite{Cosg,CosgS}. A partial classification for the third-order case was also obtained by the aforementioned authors.
The work of Bureau, \cite{Bur72, Bur87}, is also important in this respect. 
No classification results exist for the analogous discrete case and hardly any examples of 
second-order second-degree difference equations exist to date, with the notable exception of an (additive difference) equation given by Est\'evez and Clarkson \cite{Clarkson}.

A key result of this letter is a second-order second-degree equation, 
which can be considered as a $q$-analogue of  an equation in the Chazy-Cosgrove class, 
together with its Lax pair 
(i.e., isomonodromic $q$-difference problem). 
This new equation contains four free parameters, which suggests that it 
could be a $q$-difference analogue of the second-order second-degree 
differential equation that is a counterpart of the sixth Painlev\'e 
equation. 
There are several forms of a second-order second-degree equation
related to the sixth Painlev\'e 
equation that have appeared in the literature, notably
one derived by Fokas and Ablowitz \cite{FA} and another appearing in the
work of Okamoto \cite{Oka}.
Difference analogues of the Fokas-Ablowitz equation have been 
provided by Grammaticos and Ramani, \cite{RG}, but these 
difference equations were all of first-degree.  It has 
been argued by these authors that equations that are second-degree in the highest iterate cannot be viewed as \lq\lq integrable\rq\rq , however, for the new equation we establish integrability through a Lax pair in the form of an isomonodromic deformation system. 
Furthermore we show that the equation arises as a similarity reduction from an integrable partial $q$-difference equation. Through the same procedure we also construct 
higher-order second-degree equations, which form a hierarchy associated with the new equation.

\section{$q$-Difference Similarity Reduction}\label{lKdVsaac}
\setcounter{equation}{0}

Lattice equations of KdV-type were introduced and studied over the last three decades \cite{hirota:77,NQC}, 
see \cite{KDV} for a review.  These lattice equations can be formulated as partial difference equations on a lattice with step sizes that enter as parameters of the equation. 
Conventionally we think of these parameters as fixed constants. However, in agreement with the integrability of these equations, there exists the freedom to take the parameters as functions of the local lattice coordinate in each corresponding direction. 
In this paper we consider the case when the parameters depend exponentially with base $q$ on the lattice coordinates.

We work in a space ${\mathcal F}$  of functions $f$ of arbitrarily many variables $a_i$ ($i=1,\dots,M$ for any $M$) on which we define the $q$-shift operations
\[
\,_q\!T_i\,f(a_1,\ldots, a_N):=f(a_1,\ldots, q\,a_i,\ldots, a_M)\ .  
\]
For $u, v, z \in  {\mathcal F}$, we consider the following systems of nonlinear
partial $q$-difference equations:
\be\label{eq:qKdV}
\left(u-\,_q\!T_i\,_q\!T_ju\right)\left(\,_q\!T_ju-\,_q\!T_iu\right)
= (a_i^2-a_j^2) q^2 \   , 
\ee 
\be\label{eq:qmKdV}
a_j(_q\!T_jv)\,_q\!T_i\,_q\!T_jv + a_i(_q\!T_jv)v = 
a_i(_q\!T_iv)\,_q\!T_j\,_q\!T_iv + a_j(_q\!T_iv)v\   
\ee
and
\begin{eqnarray}\nonumber
&&a_i^2\left(z-\,_q\!T_iz\right)\left(\,_q\!T_jz-\,_q\!T_i\,_q\!T_jz\right)\\
\label{eq:qSKdV}&&\qquad\qquad =
a_j^2\left(z-\,_q\!T_jz\right)\left(\,_q\!T_iz-\,_q\!T_i\,_q\!T_jz\right) , 
\end{eqnarray}
where $i,j=1,\dots,M$. 
Each of these systems, (\ref{eq:qKdV}) to (\ref{eq:qSKdV}),  
represents a multi-dimensionally consistent family of partial difference equations, in the sense of \cite{NW,BS}, which implies that they constitute holonomic systems of nonlinear partial $q$-difference equations. 
Another way to formulate this property is through an underlying linear system which takes the form
\be\label{eq:Tjphi}
\,_q\!T_i^{-1}\bphi=\bM_i(k)\bphi  , 
\ee
where $\bphi=\bphi(k;\{a_j\})$ is a two-component vector-valued function and by consistency, $\,_qT_i^{-1}\,_qT_j^{-1}\bphi=\,_qT_j^{-1}\,_qT_i^{-1}\bphi$, leads to the 
set of Lax equations (for each pair of indices $i,j$)
\be\label{eq:Laxeqs}
(\,_qT_i^{-1}\bM_j)\bM_i=(\,_qT_j^{-1}\bM_i)\bM_j\  .  
\ee 
We will consider three different cases, associated respectively with equations (\ref{eq:qKdV})--(\ref{eq:qSKdV}).
To avoid proliferation of symbols we use the same symbol $\bM_i(k)$ for each of the respective Lax matrices. For specific choices of the matrices $\bM_i$ the Lax equations (\ref{eq:Laxeqs}) lead to the nonlinear equations given above. 
In the case of the $q$-lattice KdV, (\ref{eq:qKdV}), the Lax matrices $\bM_i$ are given by
\be\label{eq:MjKdV}
\bM_i(k;\{a_j\})=\frac{1}{a_i-k}\left(\begin{array}{ccc} a_i-\,_q\!T_i^{-1}u &,& 1 \\ 
k^2-a_i^2+(a_i+u)(a_i-\,_q\!T_i^{-1}u) &,& a_i+u\end{array}\right) \  .  
\ee 
In the case of the $q$-lattice mKdV, (\ref{eq:qmKdV}), the Lax matrices $\bM_i$ are given by
\be\label{eq:Mj}
\bM_i(k;\{a_j\})=\frac{1}{a_i-k}\left(\begin{array}{ccc} a_i(_q\!T_i^{-1}v)/v &,& k^2/v \\ 
\,_q\!T_i^{-1}v &,& a_i\end{array}\right) \  .  
\ee 
Finally, in the case of the $q$-lattice SKdV, (\ref{eq:qSKdV}), the Lax matrices $\bM_i$ are given by 
\be\label{eq:MjSKdV}
\bM_i(k;\{a_j\})=\frac{a_i}{a_i-k}\left(\begin{array}{ccc} 1 &,& (k^2/a_i^2)\left(z-\,_q\!T_i^{-1}z\right)^{-1} \\ 
z-\,_q\!T_i^{-1}z &,& 1\end{array}\right) \ .     
\ee 
These Lax matrices are straightforward generalizations of those 
with constant lattice parameters given in e.g. \cite{NRGO}. 

We mention that the solutions of the equations (\ref{eq:qKdV}) to (\ref{eq:qSKdV}) 
are related through discrete Miura type relations, namely
\numparts\label{eq:Miuras}
\begin{eqnarray}
&& a_i\left(z-\,_q\!T_i^{-1}z\right)=v\left(\,_q\!T_i^{-1}v\right)\quad,  \label{eq:CH} \\ 
&& s=\left(a_i-\,_q\!T_i^{-1}u\right)v-a_i\,_q\!T_i^{-1}v\quad,\label{eq:Miura} \\ 
&& \,_q\!T_i^{-1}s=a_iv-(a_i+u)\,_q\!T_i^{-1}v\  , \label{eq:Miura2} 
\end{eqnarray}
\endnumparts
where $s\in{\mathcal F}$ is an auxiliary dependent variable. From these relations, the partial $q$-difference equations (\ref{eq:qKdV}) to (\ref{eq:qSKdV}) can be derived by eliminating $s$. 

Similarity reductions of lattice equations have been considered in \cite{NP,Nijh:Dorf,NW,DIGP,NRGO} where it was shown that scaling invariance of the solution can be implemented through additional compatible constraints on the lattice equations. In the present case of (\ref{eq:qKdV})
to (\ref{eq:qSKdV}) these constraints adopt the following form \cite{FJN}
\numparts
\begin{eqnarray}
&& u(\{q^{-N}a_i\})=q^{-N}\frac{1-\ld(q^N-1)(-1)^{\sum_i\,^{q}\!\log\,a_i}}
{1+\ld(q^N-1)(-1)^{\sum_i\,^{q}\!\log\,a_i}}\,u(\{ a_j\}) , 
  \label{eq:uconstr} \\
&& v(\{q^{-N}a_i\})=\frac{1-\ld(q^N-1)(-1)^{\sum_i\,^{q}\!\log\,a_i}}{1+\mu(q^N-1)}\,v(\{ a_j\}) ,
\label{eq:vconstr} \\ 
&& z(\{q^{-N}a_i\})=q^N\frac{1-\mu(q^N-1)}{1+\mu(q^N-1)}\,z(\{ a_j\})\quad,\label{eq:zconstr} 
\end{eqnarray}
\endnumparts
where $\lambda$ and $\mu$ are constant parameters of the reduction and $N \in \mathbb{N}$  
represents a \lq\lq periodicity freedom\rq\rq.  The notation $^{q}\!\log\,x$ refers to the logarithm of $x$
with base $q$.

In order to compute the corresponding isomonodromic deformation problems associated with the 
similarity reductions we have the following constraints on the vector function of the Lax pairs. 
In the case of (\ref{eq:qKdV}) we have
\begin{eqnarray}\label{eq:phiforqKdV}
\fl\bphi(q^N\,k;\{a_j\})= \\ 
\fl\nn\left(\begin{array}{lcl}
\qquad\left(1+\ld(q^N-1)(-1)^{\sum_i\,^{q}\!\log\,a_i}\right)&,&0\\
-2\ld q \frac{q^N-1}{q-1}\left(\sum_i\,a_i\right)(-1)^{\sum_i\,^{q}\!\log\,a_i}
 &,&q^N \left(1-\ld(q^N-1)(-1)^{\sum_i\,^{q}\!\log\,a_i}\right)  \\
\end{array}\right)\bphi(k;\{q^{-N}\,a_j\}) .
\end{eqnarray}
In the case of (\ref{eq:qmKdV})
\begin{eqnarray}\label{eq:phiforqmKdV}
\fl\bphi(q^N\,k;\{a_j\})=\\
\fl\nonumber\qquad\left(\begin{array}{ll}
\left(1+\ld(q^N-1)(-1)^{\sum_i\,^{q}\!\log\,a_i}\right)&0\\
0&q^{-N}(1+\mu(q^N-1))\\
\end{array}\right)\bphi(k;\{q^{-N}\,a_j\}) .
\end{eqnarray}
In the case of (\ref{eq:qSKdV})
\begin{eqnarray}\label{eq:phiforqSKdV}
\fl\bphi(q^N\,k;\{a_j\})=\\
\fl\nonumber\qquad\left(\begin{array}{ll}
\left(1-\mu(q^N-1)\right)&0\\
0&q^{-N}(1+\mu(q^N-1))\\
\end{array}\right)\bphi(k;\{q^{-N}\,a_j\}) .
\end{eqnarray}
The similarity constraints, (\ref{eq:phiforqKdV}) to (\ref{eq:phiforqSKdV}), in conjunction with the discrete linear equations (\ref{eq:MjKdV}) to (\ref{eq:MjSKdV}) can be used to derive corresponding $q$-isomonodromic deformation problems.  
That is, (\ref{eq:phiforqKdV}) to (\ref{eq:phiforqSKdV}) lead to
$q$-difference equations in the spectral variable $k$, hence together
with the lattice equation Lax pairs we obtain $q$-isomonodromic deformation
problems for the corresponding reductions.

\paragraph{Remarks:} 
\begin{enumerate}
\item The similarity constraints above were obtained through an approach based on Jackson-type integrals, the details of which will be presented elsewhere \cite{FJN}. By construction, these constraints are compatible with the underlying lattice equations, which can be checked \textit{a posteriori} 
by an explicit calculation, presented in the appendix. 
\item In this approach, the dynamics in terms of the variables $a_i$ appear through appropriately chosen $q$-analogues of exponential functions, whereas the relevant Jackson integrals exhibit an invariance through scaling by factors $q^N$.
\item The parameters $\lambda$ and $\mu$ arise in this setting
through boundary contributions in a manner analogous to the derivation in \cite{NRGO}.
\end{enumerate}

In the remainder of this letter our aim is to implement the similarity constraint 
to obtain explicit reductions to ordinary $q$-difference equations. 
For simplicity we consider only the reduction of the $q$-mKdV equation (\ref{eq:qmKdV}), leaving
considerations of the $q$-KdV and $q$-SKdV to a future publication \cite{FJN}. 
There are two possible scenarios to derive similarity reductions of the lattice equations using the constraint (\ref{eq:vconstr}).
\paragraph{\textit{``Periodic'' similarity reduction:}} 
By fixing $M=2$ and allowing $N$ to vary, we select two lattice directions, say the variables $a_1$ and $a_2$, and consider similarity reductions with different values of $N$. This is a $q$-variant of the periodic staircase type reduction of partial difference equations on the two-dimensional lattice. 
For instance, with $N=2$ the reduction is a second-order first-degree $q$-Painlev\'e equation.
Increasing $N$ leads to $q$-difference Painlev\'e type equations of increasing order.
However, we will not pursue this route here but leave it to a subsequent publication \cite{FJN}. These reductions are reminiscent of the work \cite{Hay,SahadevanCapel,Carstea}.  
\paragraph{\textit{Multi-variable similarity reduction:}} 
The similarity constraints provide the mechanism to couple together two or more lattice directions.  
By considering the case $N=1$ we implement the similarity constraints on an extended lattice of three or more dimensions in order to obtain coupled ordinary $q$-difference equations, 
in a way that is reminiscent of the approach of \cite{NW}. 
This is considered in the next section.

\paragraph{{}} 
We have not considered the more general case of arbitrary $M, N \in \mathbb{N}$, 
which we will postpone to a future publication \cite{FJN}. 

\section{Multi-variable similarity reduction} \label{multivarsec}
\setcounter{equation}{0}
In this section we consider explicitly the $M=1$, 2 and 3 cases for $N=1$.

For simplicity we shall in what follows denote the coefficient in (\ref{eq:vconstr}) as $\gamma$, i.e. 
\be\label{eq:gamma}
\gamma=\frac{1-\ld(q-1)(-1)^{\sum_i\,^{q}\!\log\,a_i}}{1+\mu(q-1)}\quad\Rightarrow\quad
 v(\{q^{-1} a_i\})=\gamma \, v(\{a_i\})\  ,  
\ee
where $\gamma$ alternates between two values, i.e., $\,_q\!T_i^2\gamma=\gamma$. 

In contrast to the usual difference case which was explored in \cite{NW} where 
in the case of two variables we obtain a nontrivial O$\Delta$E as a reduction, 
in the $q$-difference case we have to consider at least three independent variables
to obtain a nontrivial system of O$\Delta$Es as a 
reduction.

In \cite{NW} the compatibility between the similarity constraint and the lattice 
system was established and led 
to a system of higher order difference equations in the reduction, namely equations which were on 
the level of the first Garnier system. 
In contrast to the $q=1$ work, the 3D similarity constraint here is somewhat 
simpler and leads to a second-order equation (which is of second-degree, and 
is a principal result of this letter).

\subsection*{Two-variable case}
Let us now select among the collection of variables $\{a_j\}$ two specific ones
which for simplicity we will call $a$ and $b$. 
Denote the $q$-shifts in these variables by an over-tilde $\wt{\phantom{C}}$ 
and an over-hat $\wh{\phantom{C}}$ respectively.  Equation
(\ref{eq:qmKdV}) may now be written
\be\label{qmKdVab}
b \wh{v} \wh{\wt{v}} + a \wh{v} v = a \wt{v} \wh{\wt{v}} + b v \wt{v},
\ee
where the over-tilde $\wt{\phantom{v}}$ refers to the
$q$-translation $a \mapsto q a$ and the over-hat $\wh{\phantom{v}}$ refers to the $q$-translation $b \mapsto q b$
(so if $v \equiv v(a,b)$, $\wt{v} \equiv v(q a,b)$, $\ut{v} \equiv v( q^{-1}a ,b)$, 
$\wh{v} \equiv v(a,bq)$, \ldots).

Equation (\ref{eq:gamma})
gives the constraint $v = \gamma \wh{\wt{v}}$ to impose on (\ref{qmKdVab})
(where $\wh{\wt{\gamma}} = \wt{\wt{\gamma}} = \gamma$).
This leads to the linear first-order (in that it is a two point) ordinary
difference equation
\be\label{Neq1}
\undertilde{v} = C \,\tilde{v},
\ee
where $C=\wt{\gamma} (a\gamma + b)/(a+b\gamma)$.
In the appendix the
consistency between the lattice equation (\ref{qmKdVab}) and the constraint
(\ref{Neq1}) is shown by direct computation.

\subsection*{Three-variable case}

Take three copies of the lattice mKdV equation with $a_1=a$, $a_2=b$, $a_3=c$, 
\numparts
\begin{eqnarray}\label{qmKdVabc}
b \wh{v} \wh{\wt{v}} + a \wh{v} v = a \wt{v} \wh{\wt{v}} + b v \wt{v}, \label{eq:qmKdVab} \\ 
c \ol{v} \wt{\ol{v}} + a \ol{v} v = a \wt{v} \wt{\ol{v}} + c v \wt{v}, \label{eq:qmKdVac}\\ 
c \ol{v} \wh{\ol{v}} + b \ol{v} v = b \wh{v} \wh{\ol{v}} + c v \wh{v}, \label{eq:qmKdVbc}
\end{eqnarray}
\endnumparts
where $\ol{c} = q c$, together with the constraint 
\be\label{eq:abcconstr}
v(q^{-1}a,q^{-1}b,q^{-1}c)=\gamma v(a,b,c).
\ee 
(The similarity constraint is shown by a direct computation to be compatible with the multidimensionally 
consistent system of mKdV lattice equations in the appendix.)
We now proceed to derive the 
reduced system which leads to a (higher-degree) ordinary $q$-difference 
equation in terms of one selected independent
variable, say the variable $a$. The remaining variables $b$ and $c$ 
will play the role of parameters 
in the reduced equation. Thus, 
we can derive the following system of two coupled O$\Delta$Es for $v(a,b,c)$ and 
$w(a,b,c)\equiv v(a,b,q^{-1}c)$:
\numparts\label{eq:vw}
\bea
\gamma v&=&w \frac{a\wt{\gamma}\wt{v}-b\ut{w}}{a\ut{w}-b\wt{\gamma}\wt{v}}\   ,  \label{eq:vwa} \\ 
\ut{w}&=&v \frac{aw-c\ut{v}}{a\ut{v}-cw}\   ,  \label{eq:vwb} 
\eea
\endnumparts
where $\wt{\wt{\gamma}}=\gamma$. We consider the system (\ref{eq:vwa}) and (\ref{eq:vwb})   
to constitute a $q$-Painlev\'e system with four free 
parameters. 

The system (\ref{eq:vwa}) and (\ref{eq:vwb}) 
can be reduced to a second-order second-degree ordinary difference equation 
as follows. Introduce the variables
\be\label{eq:XVW} 
X=\frac{v}{w}\quad,\quad V=\frac{\wt{v}}{v}\quad,\quad W=\frac{\wt{w}}{w}\  , 
\ee 
then from (\ref{eq:vwa}) we obtain
\be\label{eq:VW}
\wt{\gm}\frac{\wt{v}}{\ut{w}}=\wt{\gm}VX\ut{W}=\frac{a\gm X+b}{b\gm X+a}\  , 
\ee 
whereas from (\ref{eq:vwb}) we get 
\be\label{eq:VV}
W=\frac{VX}{\wt{X}}=\frac{a+q^{-1}cXV}{q^{-1}c/X+aV}\   , 
\ee 
using also the definitions (\ref{eq:XVW}). Thus, we obtain a quadratic equation for $V$ in terms 
of $X$ and $\wt{X}$ and hence also we have $W$ in terms of $X$ and $\wt{X}$. Inserting these into 
(\ref{eq:VW}) we obtain a second-order algebraic equation for $X$. Alternatively, avoiding 
the emergence of square roots, the following second-order second-degree equation for $X$ may be 
derived
\begin{eqnarray}\label{eq:Xeq}
\fl
\left[ \wt{\gm}^2\wt{X}\ut{X}-\left(\frac{a\gm X+b}{b\gm
X+a}\right)^2\right]^2    \nn \\
 = \wt{\gm}\frac{c^2}{a^2}\frac{1}{X}
\left(\frac{a\gm X+b}{b\gm X+a}\right)\left[\wt{\gm}\wt{X}(1-X\ut{X})+
q^{-1}(1-X\wt{X})\frac{a\gm X+b}{b\gm X+a}\right]  \nn \\
\qquad\quad\times \left[ q^{-1}\wt{\gm}\ut{X}(1-X\wt{X})+ (1-X\ut{X})
\frac{a\gm X+b}{b\gm X+a}\right] \  .
\end{eqnarray}
We consider this second-degree equation to be one of the main results of this letter.


We now proceed to present the Lax pair for the $q$-Painlev\'e system (\ref{eq:vwa}) and (\ref{eq:vwb}) and the second-order 
second-degree equation (\ref{eq:Xeq}).
The Lax pair is formed by considering the compatibility of two paths on the lattice:
along a `period' then in the $a$ direction and evolving in the $a$ direction
then along a `period'.
Using (\ref{eq:phiforqmKdV}) the evolution along a period is converted into a
dilation of the spectral parameter, $k$, by $q$.  
The result is the following isomonodromic $q$-difference system for the vector
$\phi(k;a)$ 
which using the results of section \ref{lKdVsaac} yields 
\numparts
\begin{eqnarray}
\phi(k;q^{-1}a)&=&\bM(k;a)\phi(k;a)\  , \\ 
\phi(qk;a)&=&\bL(k;a)\phi(k;a)\  ,  
\end{eqnarray}
\endnumparts
where 
\numparts
\be\label{eq:M}
\bM(k;a)=\frac{1}{a-k}\left(\begin{array}{cc} a\ut{v}/v & k^2/v \\ 
\ut{v} & a\end{array}\right) \  , 
\ee 
and 
\begin{equation}\label{eq:L}
\bL(k;a)= 
\frac{1}{a-k}\left(\begin{array}{cc} a\gamma v/\wt{v} & k^2/\wt{v} \\ 
q^{-1}\gamma v & q^{-1}a \end{array}\right)\,  
\left(\begin{array}{cc} b\wt{\gamma}\wt{v}/w & k^2/w \\ 
\wt{\gamma}\wt{v} & b \end{array}\right)\, 
\left(\begin{array}{cc} cw/v & k^2/v \\ 
w & c \end{array}\right)\, 
\end{equation}
\endnumparts
where we have suppressed the dependence on the variables $b$ and $c$ (which now 
play the role of parameters) and omitted the unnecessary prefactors $(b-k)^{-1}$ 
and $(c-k)^{-1}$, as well as an over factor $q^{-1}(1+\mu(q-1))$.

The consistency condition 
obtained from the two ways of expressing $\phi(qk;q^{-1}a)$ in terms of $\phi(k;a)$ is formed by the 
Lax equation  
\be\label{eq:Laxeq}
\bL(k;q^{-1}a)\bM(k;a)=\bM(qk;a)\bL(k;a)\  . 
\ee  
A gauge transformation can be obtained expressing the 
Lax matrices in terms of the variables introduced in (\ref{eq:XVW}).  Setting  
\numparts
\be \label{eq:LMa}
\mathcal{M}(k;a) = \frac{1}{a-k}\left(\begin{array}{cc} 
a/\ut{V} & k^2 \\ 1 & a\ut{V} \end{array}\right) ,
\ee
\be\label{eq:LMb}
\mathcal{L}(k;a) = \frac{1}{a-k}\left(\begin{array}{ccc} 
\wt{\gm}(ab\gm X+k^2) & & k^2(a\gm X+b)/V \\ 
q^{-1}\wt{\gm} V(a+b\gm X) & & q^{-1}(ab+k^2\gm X)\end{array}\right) 
\left(\begin{array}{cc} c/X & k^2 \\ 1/X & c\end{array}\right) \   ,  
\ee
\endnumparts
the Lax equations (\ref{eq:Laxeq}) (replacing $\bL$ and $\bM$ by $\mathcal{L}$ and $\mathcal{M}$ respectively) 
yield a set of relations equivalent to the following two equations: 
\begin{eqnarray}\label{eq:laxconds}
&& \wt{\gm}V\ut{V}\ut{X}=\frac{a\gm X+b}{a+b\gm X}  \   , \label{eq:laxcondsa} \\ 
&& aV^2+q^{-1}c\left( \frac{1}{X}-\wt{X}\right) V-a\frac{\wt{X}}{X}=0\   , \label{eq:Laxcondsb} 
\end{eqnarray} 
using also $\wt{\gm}=\ut{\gm}$.  This set follows directly from (\ref{eq:VW}) and (\ref{eq:VV}). Thus 
(\ref{eq:LMa}) and (\ref{eq:LMb}) 
form a $q$-isomonodromic Lax pair for the second-degree equation (\ref{eq:Xeq}).

\subsection*{Four-variable case:}

Suppose we have $4$ variables  $a_i$, $i=1,\dots, 4$.  Select $a=a_1$ to be the 
independent variable after reduction. 
Introduce the dependent variables $w_{j-2}=\,_q\!T_j^{-1}v$, 
$j=3,4$. Then directly from the $q$-lattice mKdV equation (\ref{eq:qmKdV}) we have the set 
of equations
\be\label{eq:multi-qmKdV}
\underaccent{\wtilde}{w}_j
=v\frac{aw_j-a_{j+2}\ut{v}}{a\ut{v}-a_{j+2}w_j}\quad,\quad j=1,2\  , 
\ee
where as before the tilde denotes a $q$-shift in the variable $a$. 
At the same time the multiply shifted
object $_q\!T_3^{-1} \,_q\!T_4^{-1}\ut{v}$ can be expressed in a unique way (due to the CAC property) 
in terms of $\ut{v}$ and $\,_q\!T_j^{-1}v=w_{j-2}$, $j=3,4$, by iterating the relevant copies of the 
$q$-lattice mKdV equation in the variables $a_j$, $j\neq 2$, leading to an expression of the form
$_q\!T_3^{-1} \,_q\!T_4^{-1}\ut{v}=:F(\ut{v},w_1,w_{2})$, where $F$
is easily obtained explicitly. 
Imposing the similarity constraint (\ref{eq:gamma}) 
we obtain $\wt{\gm}\,_q\!T_2v=F(\ut{v},w_1,w_{2})$
and inserting this expression into the $q$-lattice mKdV (\ref{eq:qmKdV}) with $i=1,j=2$ we obtain 
\be\label{eq:vvF}
\left( a+\frac{a_2}{\gm} \frac{a_3 w_2-a_4 w_1}{a_3 w_1 -a_4 w_2}\right)
\left(a_2\wt{\gm}^{-1}F(\ut{v},w_1,w_{2})-a\wt{v}\right)=
(a_2^2-a^2)v\wt{v}\  . 
\ee 
With the explicit form of $F(\ut{v},w_1,w_{2})$ equation (\ref{eq:vvF}) reads
\numparts
\begin{eqnarray}\label{eq:4dsysta}
\fl
(a_2^2-a^2)\gm\wt{\gm}\wt{v}(a_3w_1-a_4w_2)
\left[ a(a_3^2-a_4^2)\ut{v}+a_3(a_4^2-a^2)w_1+a_4(a^2-a_3^2)w_2\right] \nn \\
=[(a_2a_3-\gm aa_4)w_2+(\gm aa_3-a_2a_4)w_1]\,\left[ a(a_3^2-a_4^2)(a_2 w_1w_2-\wt{\gm}a\wt{v}\ut{v}) \right. 
\nn \\
\left. +\left(a_2a_4(a^2-a_3^2)\ut{v}-aa_3(a_4^2-a^2)\wt{\gm}\wt{v}\right)w_1 +
\left(a_2a_3(a_4^2-a^2)\ut{v}-\wt{\gm}aa_4(a^2-a_3^2)\wt{v}\right)w_2 \right] \  , \nn \\ 
\eea 
and this is supplemented by the two equations
\bea
&& avw_1+a_3w_1\underaccent{\wtilde}{w}_1=a_3v\ut{v}+a\ut{v}\underaccent{\wtilde}{w}_1\   , \label{eq:4dsystb} \\ 
&& avw_2+a_4w_2\underaccent{\wtilde}{w}_2=a_4v\ut{v}+a\ut{v}\underaccent{\wtilde}{w}_2\   , \label{eq:4dsystc} 
\eea
\endnumparts
which is equivalent to a five-point 
(fourth-order) $q$-difference equation in terms of 
$v$ alone, containing five free parameters: $a_2$, $a_3$, $a_4$, $\ld$ and $\mu$ (inside $\gm$ and $\wt{\gm}$). This would be an algebraic equation, 
so we proceed as follows in order to derive a higher-degree 
$q$-difference system. Introduce the variables
\be\label{eq:XW}
X_i=\frac{v}{w_i}\quad,\quad  W_i=\frac{\wt{w}_i}{w_i}\quad,\quad i=1,2\  , 
\ee 
while retaining the variable $V=\wt{v}/v$ as before. By definition we have
\be\label{eq:XWV}
\frac{V}{W_i}=\frac{\wt{X}_i}{X_i}\quad,\quad i=1,2  , 
\ee
and from (\ref{eq:4dsystb}), (\ref{eq:4dsystc}) we obtain
\be\label{eq:WXV}
W_i=\frac{qa+a_{i+2}VX_i}{qaV+a_{i+2}/X_i}=\frac{VX_i}{\wt{X}_i}\quad,\quad i=1,2\ , 
\ee 
whilst from (\ref{eq:4dsysta}) we get
\begin{eqnarray}\label{eq:WVX} 
\fl
(a_2^2-a^2)\gm\wt{\gm}V\left(\frac{a_3}{X_1}-\frac{a_4}{X_2}\right)\left( a\frac{a_3^2-a_4^2}{\ut{V}}+
a_3\frac{a_4^2-a^2}{X_1}+a_4\frac{a^2-a_3^2}{X_2}\right)  \nn \\
 =\left(\frac{a_2a_3-\gm aa_4}{X_2}+\frac{\gm aa_3-a_2a_4}{X_1}\right)\,\left[ a(a_3^2-a_4^2)(\frac{a_2}{X_1X_2}
-\wt{\gm}a\frac{V}{\ut{V}})  \right. \nn \\
\left. +\left(a_2a_4\frac{a^2-a_3^2}{\ut{V}}-aa_3(a_4^2-a^2)\wt{\gm}V\right)\frac{1}{X_1} +
\left(a_2a_3\frac{a_4^2-a^2}{\ut{V}}-\wt{\gm}aa_4(a^2-a_3^2)V\right)\frac{1}{X_2} \right] \  . \nn \\ 
\end{eqnarray}
{}From (\ref{eq:WXV}) we obtain the set of quadratic equations for $V$
\be
qa\frac{X_i}{\wt{X}_i}V^2+a_{i+2}\left(\frac{1}{\wt{X}_i}-X_i\right)V - qa=0\quad,\quad i=1,2\  , 
\ee 
from which by eliminating $V$ we obtain
\bea\label{eq:YX}
\fl
\left[a_3(1-X_1\wt{X}_1)X_2-a_4(1-X_2\wt{X}_2)X_1\right]\left[a_3(1-X_1\wt{X}_1)\wt{X}_2-a_4(1-X_2\wt{X}_2)\wt{X}_1\right] \nn \\ 
= q^2a^2 (X_1\wt{X}_2-X_2\wt{X}_1)^2\   . 
\eea 
Furthermore, solving $V$ from the quadratic system as
\be\label{eq:V}
V=qa\frac{X_2\wt{X}_1-X_1\wt{X}_2}{a_3(1-X_1\wt{X}_1)X_2-a_4(1-X_2\wt{X}_2)X_1}\  ,  
\ee 
and inserting this into (\ref{eq:WVX}) 
we obtain a second-order equation in both $X_1$, $X_2$ coupled to the equation 
(\ref{eq:YX}) which is first order in both $X_1$, $X_2$. It is this coupled system of two equations in $X_1$, $X_2$ which 
forms our higher order generalisation of (\ref{eq:Xeq}). 
The system of (\ref{eq:WVX}) and (\ref{eq:YX}) with (\ref{eq:V})
constitutes a third-order system with five parameters.

The derivation of the Lax pair follows the same approach as that for the
three-variable case
(with an extra factor in $\mathcal{L}$ due to the additional lattice direction).
We omit details here, which we intend to publish in the future \cite{FJN}.

\subsection*{Beyond the four-variable case:} 

It is straight-forward to give the form of the full hierarchy, however due to lack of space we postpone this until a later publication \cite{FJN}.

\section{Conclusion and discussion}\label{limsanddeg}
\setcounter{equation}{0}

In this letter we have presented the results of 
a scheme to derive partial $q$-difference equations 
of KdV type and consistent symmetries of the equations 
and demonstrated how it can be implemented.  
Lax matrices follow from this approach. A notable result is the 
derivation of the higher-degree equation (\ref{eq:Xeq}), showing that 
the scheme presented here allows for the derivation of new results
within the field of discrete integrable systems.  

The first-, second- and third-order members of the $N=1$ hierarchy have been presented.
The scheme continues to give successively higher-order
equations by considering successively higher dimensions of the original lattice equation.  
One may ask the natural question
as to whether this gives an `interpolating' hierarchy which,
contrary to the usual cases, increases the order and number of parameters
of the equations
by one in each step, rather than a two step increase.  A further natural
question connected with this hierarchy is its
relation to the $q$-Garnier systems of Sakai \cite{GarnSakai}.

We will present full details of the 
scheme from which the lattice equations (\ref{eq:qKdV})
to (\ref{eq:qSKdV}) and their associated constraints follow in a future publication \cite{FJN}.
There we will consider the most general case of symmetry reductions 
(arbitrary $N \in \mathbb{N}$) of all three lattice equations.

We also intend 
to return in a future publication to the question of limits and degeneracies of the 
equations presented in this paper.  These include the 
$q\rightarrow 1$ continuum limit,
the $q\rightarrow 1$ discrete limit 
and the $q\rightarrow 0$ crystal or ultradiscrete limit. 

\section*{Acknowledgements} 

During the writing of this letter C.M. Field has been supported by
the Australian Research Council Discovery Project Grant $\#$DP0664624
and by the Netherlands Organization
for Scientific Research (NWO) in the VIDI-project ``Symmetry and
modularity in exactly solvable models''.
N. Joshi is also supported by 
the Australian Research Council Discovery Project Grant $\#$DP0664624.
Part of the work presented here was performed whilst N. Joshi 
was visiting the University of Leeds.

\section{Appendix}

In this appendix we present the result of explicit calculations showing the consistency
of the lattice equations and constraints.

\subsection*{Two-variable consistency}

We shall check the consistency between the lattice equation (\ref{qmKdVab}) and the 
constraint $v= \gamma \wh{\wt{v}}$ by direct computation. 
This computation is illustrated in the following diagram: 
\vspace{.2cm}
\begin{center}

\setlength{\unitlength}{0.00043489in}
\begingroup\makeatletter\ifx\SetFigFont\undefined%
\gdef\SetFigFont#1#2#3#4#5{%
  \reset@font\fontsize{#1}{#2pt}%
  \fontfamily{#3}\fontseries{#4}\fontshape{#5}%
  \selectfont}%
\fi\endgroup%
{\renewcommand{\dashlinestretch}{30}
\begin{picture}(5245,4530)(0,-10)
\put(-1010,2205){.}
\put(-1010,3985){.}
\put(-1010,380){.}

\put(6235,2205){.}
\put(6235,3985){.}
\put(6235,380){.}

\put(765,3985){.}
\put(4410,380){.}

\put(2565,2205){\circle*{303}}
\put(4410,2205){\circle*{303}}
\put(765,2205){\circle{303}}
\put(2365,3885){{\Large$\times$}}

\put(4410,4005){\circle{303}}
\put(615,325){{\large$\otimes$}}
\put(2610,439){\circle{303}}
\thicklines
\dashline{90.000}(4410,4005)(2565,4005)(765,2205)
	(765,405)(2610,405)(4410,2205)(4410,4005)
\drawline(2565,2205)(4410,2205)
\put(2385,1725){\makebox(0,0)[lb]{\Large $v_0$}}
\put(4410,1770){\makebox(0,0)[lb]{\Large $v_1$}}
\put(4815,3980){\makebox(0,0)[lb]{\Large $v_{12}$}}
\put(2340,4305){\makebox(0,0)[lb]{\Large $v_2$}}
\put(0,2205){\makebox(0,0)[lb]{\Large $v_{-1}$}}
\put(295,-15){\makebox(0,0)[lb]{\Large $v_{-1,-2}$}}
\put(2475,-75){\makebox(0,0)[lb]{\Large $v_{-2}$}}
\end{picture}
}

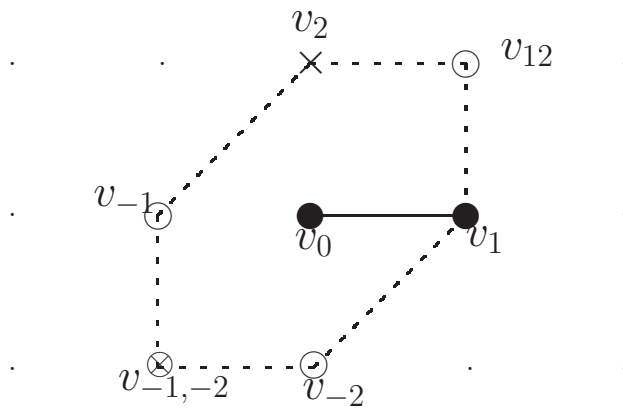
Fig 1.  Consistency on the 2D lattice.  
\end{center}
\noindent
Assuming the values $v_0$, $v_1$ as indicated in Fig 1 are given, 
we compute successively $v_{12}$,$v_2$ etc., 
where the subscripts refer to the shifts in the lattice variables $a$,$b$ respectively, as is evident 
from Fig 1.
Points other than $v_0$ and $v_1$ are computed using either 
the lattice equation (indicated by $\times$) or by using the similarity constraint (indicated by 
$\bigcirc$). The value $v_{-1,-2}$ is the first point which can be calculated in two different ways 
(hence indicated in the diagram by $\otimes$). Without making any particular assumptions on how 
$\gamma$ depends on $a$ and $b$, a straightforward calculation shows that the two ways of computing 
$v_{-1,-2}$ are indeed the same, for any choice of initial data $v_0$ and $v_1$, provided that $\gamma$ 
obeys the relation
\be\label{eq:gammarel}
\left(\frac{a+b\gamma}{b+a\gamma}\right)^{\!\wh{\wt{\phantom{a}}}}
\left(\frac{a+b\gamma}{b+a\gamma}\right)^{-1}=\frac{\wt{\gamma}}{\wh{\gamma}}\   . 
\ee 
A particular solution of this relation is
\be\label{eq:gammasol} 
\wh{\wt{\gamma}}=\gamma\quad\Leftrightarrow\quad \wh{\gamma}=\wt{\gamma}\  , 
\ee 
and hence $\wt{\wt{\gamma}}=\gamma$ implying that $\gamma$ is an alternating ``constant'' which is in 
accordance with the value given in (\ref{eq:vconstr}). The reduced equation 
in this case is (\ref{Neq1}),
which can be readily integrated. 

More generally, equation (\ref{eq:gammarel}) can be resolved by setting 
\be\label{eq:nurels} \frac{a+b\gamma}{b+a\gamma}=\frac{\wt{\nu}}{\wh{\nu}}\quad,\quad 
\gamma=\frac{\wh{\wt{\nu}}}{\nu}\  ,  \ee 
leading to the consequence that $\nu$ has to solve the $q$-lattice mKdV (\ref{qmKdVab}). 
In principle we could take for $\nu$ any solution of the reduced equation (\ref{Neq1}) 
and use this to parametrise the reduced equation for $v$ via the relations (\ref{eq:nurels}).   
In any event, we see that the two-variable case does not lead to interesting nonlinear equations.

\subsection*{Three-variable consistency}
In this 
case the consistency diagram is as follows: 
\vspace{.2cm} 

\begin{center}

\setlength{\unitlength}{0.00057489in}
\begingroup\makeatletter\ifx\SetFigFont\undefined%
\gdef\SetFigFont#1#2#3#4#5{%
  \reset@font\fontsize{#1}{#2pt}%
  \fontfamily{#3}\fontseries{#4}\fontshape{#5}%
  \selectfont}%
\fi\endgroup%
{\renewcommand{\dashlinestretch}{30}
\begin{picture}(6666,3630)(0,-10)
\put(3465,405){\circle{180}}
\put(5265,1800){\circle{180}}
\put(2430,1760){$\otimes$}
\put(1800,1665){\circle*{180}}
\put(675,405){\circle{180}}
\put(4200,1580){$\times$}
\put(6210,3060){\circle*{180}}
\put(3510,3105){\circle*{180}}
\thicklines
\drawline(720,405)(3465,405)
\drawline(3510,405)(5220,1800)
\drawline(3465,3105)(6210,3105)
\drawline(1778,1682)(3488,3077)
\dashline{90.000}(3465,3105)(3465,405)
\dashline{90.000}(2610,1800)(5175,1800)
\dashline{90.000}(2565,1800)(720,450)
\dashline{90.000}(1800,1665)(4365,1665)
\dashline{90.000}(6187,3017)(4342,1667)
\put(3240,3465){\makebox(0,0)[lb]{\Large $v_0$}}
\put(6390,3375){\makebox(0,0)[lb]{\Large $v_1$}}
\put(1080,1710){\makebox(0,0)[lb]{\Large $v_2$}}
\put(3375,0){\makebox(0,0)[lb]{\Large $v_{-3}$}}
\put(4000,1300){\makebox(0,0)[lb]{\Large $v_{1,2}$}}
\put(5500,1750){\makebox(0,0)[lb]{\Large $v_{-2,-3}$}}
\put(0,0){\makebox(0,0)[lb]{\Large $v_{-1,-3}$}}
\put(2565,1925){\makebox(0,0)[lb]{\Large $v_{-1,-2,-3}$}}
\end{picture}
}

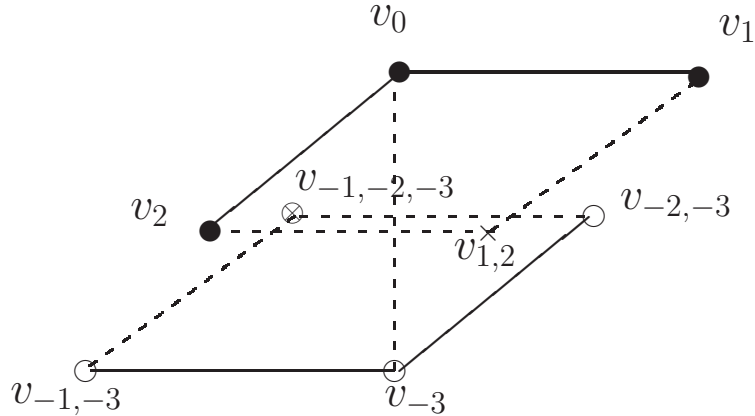
Fig 2. Consistency on the 3D lattice.
\end{center}
\vspace{.2cm}

\noindent 
A similar notation as the previous case is used as is evident from Fig 2.
The initial data $v_0$, $v_1$ and $v_2$ are given, and the indicated values on the vertices are 
computed either by using one of the lattice equations (\ref{eq:qmKdVab}) to (\ref{eq:qmKdVbc}) 
or the similarity 
constraint (\ref{eq:abcconstr}) over the diagonal. Thus, $v_{1,2}$ is obtained from (\ref{eq:qmKdVab}) 
yielding 
$$ v_{1,2}=v_0\,\frac{av_2-bv_1}{av_1-bv_2}\  ,   $$
whilst from the similarity constraint we obtain 
$$v_{-1,-3}=\gm_2 v_2 \quad,\quad v_{-3}=\gamma_{1,2}v_{1,2}\quad,\quad 
v_{-2,-3}= \gm_1 v_1\  , $$
assuming that $\gm$ shifts along the lattice, indicated by the indices, 
and finally the value of $v_{-1,-2,-3}$ can be computed in 
two different ways, leading to
$$v_{-1,-2,-3}=\gm_0 v_0=\frac{av_{-2,-3}-bv_{-1,-3}}{av_{-1,-3}-bv_{-2,-3}}v_{-3}=
\frac{a\gm_1v_1-b\gm_2v_2}{a\gm_2v_2-b\gm_1v_1}\gm_{1,2}v_0 \frac{av_2-bv_1}{av_1-bv_2}\  , $$ 
leading quadratic identity in $v_1$ and $v_2$. Assuming that the latter must hold identically, 
and thus setting all coefficients equal to zero, we obtain the following conditions on $\gm$:
$$ \gm_{1,2,3}=\gm_1=\gm_2=\gm_3\quad , $$ 
from which we conclude that $\gm$ is an alternating ``constant'', for instance 
\be\label{form:gamma}
\gm=\alpha\,\beta^{(-1)^{n+m+\dots}}\quad  \quad (\alpha,\beta\ \ {\rm constants})  \  
\ee 
and this leads to the conditions
from which it is easily deduced that the form (\ref{eq:gamma}) of $\gamma$ satisfies these 
conditions.


\section*{References}
\begin{thebibliography}{99}

\bibitem{BS}
Bobenko A. and Suris Y.:
\newblock Integrable systems on quad-graphs,
{\em Int. Math. Res. Not.} {\bf 11}  573--611 (2002)

\bibitem{Birkhoff}
Birkhoff G. D.:
\newblock The generalized Riemann problem for linear differential equations and 
the allied problems for linear difference and $q$-difference equations, 
{\em Proc. Am. Acad. Sci.} 
{\bf 49} 521--568 (1913) 

\bibitem{Bur72}
Bureau F. J.: 
\newblock
\'Equations diff\'erentielles du second ordre en 
$Y$ et du second degr\'e en $\ddot Y$ dont l'int\'egrale g\'en\'erale est \'a points critiques fixes,
{\em Ann. Mat. Pura Appl.} {\bf 91} (4) 163--281 (1972)

\bibitem{Bur87}
Bureau F. J.: 
\newblock
Sur des syst\`emes diff\'erentiels non lin\'eaires du troisi\`eme ordre et les \'equations diff\'erentielles non lin\'eaires associ\'ees, 
{\em Acad. Roy. Belg. Bull. Cl. Sci.} {\bf 73} (5) no. 6-9, 335--353 (1987)
 

\bibitem{Ca:complete}
Capel H. W., Nijhoff F. W. and Papageorgiou V. G.: 
\newblock Complete integrability of Lagrangian mappings and lattices of KdV
  type
\newblock {\em Phys. Lett. A} {\bf 155} 377--387 (1991)

\bibitem{Chazy9}
Chazy J.:
Sur les \'equations diff\'erentielles du second ordre \`a points
critiques fixes 
{\em Comptes Rendus de l'Acad\'emie des Sciences, Paris} {\bf 148} 
1381-1384 (1909)

\bibitem{Chazy11}
Chazy J.:
Sur les \'equations diff\'erentielles du troisi\`eme ordre et
ordre sup\'erieur dont
l'int\'egrale g\'en\'erale a ses points critiques fixes,
{\em Acta Math.} {\bf 34} 317--385 (1911)


\bibitem{Cosg}
Cosgrove C.: Chazy's second-degree Painlev\'e equations, 
{\em J. Phys. A: Math. Gen.} {\bf 39} 11955-11971 (2006) 

\bibitem{CosgS}
Cosgrove C.M., Scoufis G.: 
Painlev\'e Classification of a Class 
of Differential Equations of the Second Order and Second Degree, 
{\it Stud. Appl. Math.} 88:25--87 (1993)

\bibitem{Clarkson}
Est\'evez P. G. and Clarkson P. A.: 
Discrete equations and the singular manifold method,
In: D. Levi and O. Ragnisco (eds), {\em SIDE III---symmetries and integrability of difference equations}
CRM Proceedings and Lecture Notes, 25. 
pp 139--146,
American Mathematical Society, Providence, RI, (2000)

\bibitem{FJN}
Field C.M., Joshi N. and Nijhoff F.W.: in preparation.

\bibitem{FA}
Fokas A.S. and Ablowitz M.J.: 
A unified approach to transformations and elementary solutions of
Painlev\'e equations, {\it J. Math. Phys.} {\bf 23} 2033--2042 (1982)

\bibitem{FIK}
Fokas A.S., Its A.R. and Kitaev A.V.,
Discrete Painlev\'e Equations and their Appearance in Quantum Gravity,
{\em Comm. Math. Phys.} {\bf 142} 313--344 (1991) 


\bibitem{GNR}
Grammaticos B., Nijhoff F. W. and Ramani A.: Discrete Painlev\'e equations, 
In: R. Conte (ed), {\em The Painlev\'e Property: One Century Later}, pp 413--516,
Springer-Verlag, New York (1999)

\bibitem{Garn}
Garnier R.: Sur des \'equations diff\'erentielles du troisi\`eme 
ordre dont l'int\'egrale g\'en\'erale est uniforme et sur une 
classe d'\'equations nouvelles d'ordre sup\'erieur, {\it Ann. 
\'Ecol. Norm. Sup.} {\bf 29} 1--126 (1912) 

\bibitem{GR}
Grammaticos B. and Ramani A.: 
Discrete Painlev\'e equations: A review, In: 
B. Grammaticos, Y. Kossmann-Schwarzbach and 
T. Tamizhmani (eds)
{\em Discrete Integrable Systems}, Lect. Notes Phys. {\bf 644}, pp. 245--321, Springer Verlag, 2004),  
(2004)

\bibitem{Carstea}
Grammaticos B., Ramani A., Satsuma J., Willox R. and Carstea A.: 
Reductions 
of integrable lattices, 
{\em J. Nonl. Math. Phys.} {\bf12} Suppl. 363--371 (2005)

\bibitem{Hay}
Hay M., Hietarinta J., Joshi N. and Nijhoff F. W.: 
A Lax pair for a lattice modified 
KdV equation, reductions to $q$-Painlev\'e equations and associated Lax pairs, 
{\em J. Phys. A: Math. Theor.} {\bf 40} F61--F73 (2007) 

\bibitem{hirota:77}
Hirota R.:
\newblock Nonlinear Partial Difference Equations I. A difference analogue of the Korteweg-de Vries equation
\newblock {\em J. Phys. Soc. Japan} {\bf 43} (4) 1424-1433 (1977)

\bibitem{hydon} 
Hydon P. and Rasin O.:
\newblock Symmetries of integrable difference equations on the quad-graph, 
\newblock {\em Stud. Appl. Math.} {\bf 119} (3) 253-269 (2007)

\bibitem{JimSakai}
Jimbo M. and Sakai H.: 
A $q$-analog of the sixth Painlev\'e equation, 
{\em Lett Math. Phys.} {\bf 38} 145--154 (1996) 


\bibitem{Kajiwara}
Kajiwara K., Noumi M. and Yamada Y.: 
A study on the fourth $q$-Painlev\'e equation, 
\newblock {\em J. Phys. A} {\bf 34} 8563--8581 (2001)   


\bibitem{Kajiwara2}
Kajiwara K., Noumi M. and Yamada Y.: 
Discrete dynamical systems with $W(A_{m-1}^{(1)}\times 
A_{n-1}^{(1)})$ symmetry, 
\newblock {\em Lett. Math. Phys.} {\bf 60} 211-219 (2002)   

\bibitem{Noumi}
Kajiwara K., Noumi M. and Yamada Y.:  
$q$-Painlev\'e systems arising from $q$-KP hierarchy, 
\newblock {\em Lett. Math. Phys.} {\bf 62} 259--268 (2002)

\bibitem{Nijh:Dorf}
Nijhoff F. W.: 
On some ``Schwarzian'' Equations and their discrete
analogues, 
In: A.S. Fokas and I.M. Gel'fand (eds),
{\em Algebraic Aspects of Integrable Systems: In memory of Irene Dorfman}, pp 237--260. Birkh\"auser Verlag (1996)

\bibitem{DIGP}
Nijhoff F. W.: 
Discrete Painlev\'e equations and symmetry reduction on the lattice, In: 
A.I. bobenko and R. Seiler (eds)
\textit{Discrete Integrable geometry and Physics}, pp. 209--243, Oxford 
Univ. Press,  (1999) 

\bibitem{KDV}
Nijhoff F. W. and Capel H. W.: 
The discrete Korteweg-de Vries equation, 
{\em Acta Appl.
Math.} {\bf 39} 133--158 (1995)

\bibitem{NP}
Nijhoff F. W. and Papageorgiou V.G.:
Similarity reductions of integrable
lattices and discrete analogues of the Painlev\'e II equation,
{\em Phys. Lett. A} {\bf 153} 337--344 (1991)

\bibitem{NQC} 
Nijhoff F. W., Quispel G.R.W. and Capel H.W.: 
Direct linearization of nonlinear difference-difference equations, 
{\em Phys. Lett. A} {\bf 97} 125--128 (1983)

 
\bibitem{NRGO} 
Nijhoff F. W., Ramani A., Grammaticos B. and Ohta Y.: 
On discrete Painlev\'e equations associated with the lattice KdV systems 
and the Painlev\'e VI equation, 
{\em Stud. Appl. Math.} {\bf 106} 261--314 (2001)

\bibitem{NW}
Nijhoff F. W. and Walker A. J.: 
The discrete and continuous Painlev\'e VI hierarchy and the Garnier systems, 
{\em Glasgow Math. J.} {\bf 43A} 109--123 (2001)

\bibitem{Oka}
Okamoto K.:
Studies on the Painlev\'e equations. 1. 6th Painlev\'e equation PVI.
{\em Ann. Math.} {\bf 146} 337--381 (1987)


\bibitem{PNGR}
Papageorgiou V. G., Nijhoff F. W., Grammaticos B. and Ramani A.:
Isomonodromic deformation problems for discrete analogues of
Painlev\'e equations, 
{\em Phys. Lett. A} {\bf 164} 57--64. (1992)

\bibitem{RG} 
Ramani A. and Grammaticos B.: 
On the discrete form of the Fokas-Ablowitz equation, 
{\em Chaos, Solitons and Fractals} {\bf 24} 1331-1335 (2005)


\bibitem{RGH}
Ramani A., Grammaticos B. and Hietarinta J.: 
Discrete versions of the
Painlev\'e equations,  
{\em Phys. Rev. Lett.} {\bf 67} 1829--1832 (1991)

\bibitem{SahadevanCapel}
Sahadevan R. and Capel H. W.: 
Complete integrability and singularity confinement
of nonautonomous modified Korteweg-de Vries and sine Gordon mappings, 
{\em Physica A} {\bf 330} 373--390 (2003)


\bibitem{Sakai}
Sakai H.: 
Rational surfaces associated with affine root systems and geometry of the 
Painlev\'e equations, 
{\em Comm. Math. Phys.} {\bf 220} 165--229 (2001) 


\bibitem{GarnSakai}
Sakai H.: 
A $q$-analog of the Garnier system, 
{\em Funkcial. Ekvac.} {\bf 48} 273--297 (2005)
 
\bibitem{qSakai}
Sakai H.: 
Lax form of the $q$-Painlev\'e equation associated with the $A_2^{(1)}$ 
surface, 
{\em J. Phys. A: Math. Gen.} {\bf 39} 12203--12210 (2006)



\end{thebibliography}
\end{document}